# Second law of thermodynamics:
# Spontaneous cold-to-hot heat transfer in a nonchaotic medium

Yu Qiao[1,2,*], Zhaoru Shang[1]

[1] *Program of Materials Science and Engineering, University of California – San Diego, La Jolla, CA 92093, U.S.A.*

[2] *Department of Structural Engineering, University of California – San Diego, La Jolla, CA 92093-0085, U.S.A.*

[*] Corresponding author (email: yqiao@ucsd.edu)

**Abstract.** It has long been known that, fundamentally different from a large body of rarefied gas, when a Knudsen gas is immersed in a thermal bath, it may never reach thermal equilibrium. The root cause is nonchaoticity: as the particle-particle collisions are sparse, the particle trajectories tend to be independent of each other. Usually, this counterintuitive phenomenon is studied through kinetic theory and is not considered a thermodynamic problem. In current research, we show that if incorporated in a compound setup, such an intrinsically nonequilibrium behavior has nontrivial consequences and cannot circumvent thermodynamics: cold-to-hot heat transfer may happen spontaneously, either continuously (with an energy barrier) or cyclically (with time-dependent entropy barriers). It allows for production of useful work by absorbing heat from a single thermal reservoir without any other effect. As the system obeys the first law of thermodynamics, it breaks the boundaries of the second law of thermodynamics.

## 1. Introduction

An ideal gas (e.g., a classical rarefied gas) is referred to as a Knudsen gas when the mean free path of the gas particles ($\lambda_F$) is larger than the characteristic size of the gas container ($D$) [1], as depicted in Figure 1(a-c). When $D \ll \lambda_F$, compared to the particle-wall collisions, the particle-particle collisions are sparse, so that their effect is secondary. In the interior of the container, the particle trajectories tend to be nonchaotic [2]. As a result, the conventional thermodynamic analysis methods, such as the Bhatnagar-Gross-Krook (BGK) model for the Boltzmann equation [3], may not be applicable. Usually, a Knudsen gas is studied through kinetic theory rather than statistical mechanics. It could have unique fluid dynamic properties (e.g., the Knudsen paradox [4]) and unusual thermal properties (e.g., the Knudsen effect [5]). In mesoscopic physics, similar nonequilibrium phenomena are also observed, e.g., the ballistic transport of charge carriers [6].

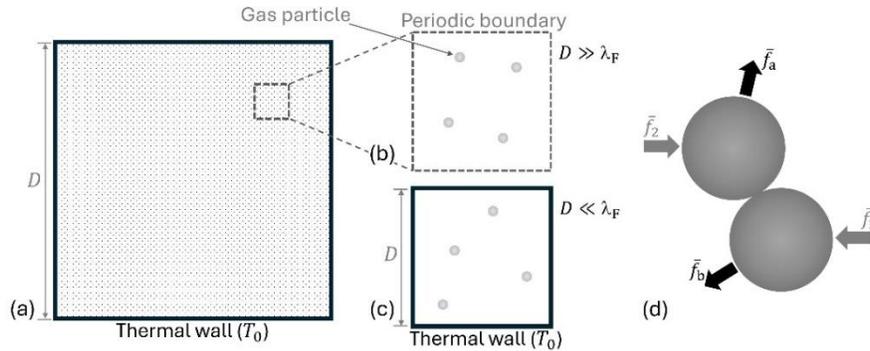

**Figure 1.** **(a)** A chaotic ideal gas (e.g., a large body of rarefied gas) immersed in a thermal bath. The container size ($D$) is much larger than the mean free path of particle-particle collision ($\lambda_F$). The container boundaries are thermal walls at a constant temperature $T_0$. **(b)** A small volume of the ideal gas in panel (a) may be modeled by using periodic boundary condition. **(c)** A Knudsen gas immersed in a thermal bath. The container boundaries are thermal walls, fundamentally different from panel (b). As $D \ll \lambda_F$, in the interior of the container, the particle trajectories tend to be independent of each other. Consequently, the system cannot relax to thermal equilibrium, i.e., $T < T_0$. **(d)** Boltzmann's H-theorem is based on the hypothesis of molecular chaos, which requires extensive particle-particle collisions throughout the system. In a Knudsen gas, however, particle-particle collisions are sparse.

People are well aware that when a Knudsen gas is immersed in a thermal bath, it may never reach thermodynamic equilibrium [7,8]. Compare Figure 1(c) to Figure 1(b). In Figure 1(b), the particle speed ($v$) follows the Maxwell-Boltzmann distribution. In Figure 1(c), with a different boundary condition, the distribution of $v$ is non-Boltzmannian. At the steady state, the effective gas-phase kinetic temperature ($T \propto \overline{K}/k_B$) of a Knudsen gas can be significantly lower than the container-wall temperature ($T_0$) [9-11], where $\overline{K}$ is the average particle kinetic energy, and $k_B$ is the Boltzmann constant. This is because, without extensive particle-particle collisions, the fast gas particles tend to rapidly move across the container and release heat to the container walls, while the slow gas particles tend to stay long in the interior. At any moment, compared to thermal equilibrium, it is more likely to find slow particles in the Knudsen gas and less likely to find fast particles.

One perspective to understand the counterintuitive phenomenon of $T < T_0$ is to examine the extreme case wherein the container contains only one gas particle. Since it is impossible for the particle to encounter another particle, the effective mean free path of particle-particle collision $\lambda_F \rightarrow \infty$. The mean free path of particle-wall collision may be assessed as $\lambda_w = \pi D/4$ [12]. Because $\lambda_w \ll \lambda_F$, the system behavior is dominated by the particle-wall collisions. When the container is immersed in a thermal bath, the probability density ($\rho_L$) of finding the particle at a speed $v$ is proportional to $p_w(v)/v$, where $p_w(v)$ is the Maxwell-Boltzmann distribution of particle speed at the container boundaries (the thermal walls). In the ideal-case scenario, with the non-Boltzmann $\rho_L$, $T = T_0/2$ in a two-dimensional Knudsen gas and $T = 2T_0/3$ in a three-dimensional Knudsen gas (see Section A1 in the Appendix for the detailed discussion).

In the past, Knudsen gases were often investigated in simple setups, without external force fields (energy barriers) or time-dependent hurdles of particle-particle collisions (entropy barriers). Under this condition, the intrinsically nonequilibrium steady state ($T < T_0$) was viewed as "trivial", partly because $T$ can be theoretically calculated but cannot be directly measured. If a temperature sensor is used, the measurement result at the sensor-gas interface would be the container-wall temperature ($T_0$), compatible with the zeroth law of thermodynamics. In other words, a Knudsen gas does not behave as a heat sink in an otherwise uniform temperature field.

However, recent research on spontaneously nonequilibrium dimension (SND) [13,14] raises a critical question: what if a Knudsen gas is combined with an energy barrier or a time-dependent entropy barrier? The theoretical and numerical analyses in [13,15] demonstrated that, with locally nonchaotic energy barriers, there may be macroscopic non-thermodynamic systems having nontrivial energy properties (Figure 2). Useful work can be produced by absorbing heat from a single thermal reservoir without any other effect. The numerical and experimental study in [14] drew a similar conclusion for locally nonchaotic entropy barriers, as entropy can spontaneously decrease without any energetic penalty.

The concern of the Knudsen gas is reinforced by the fact that Boltzmann's H-theorem [3] is inapplicable if the hypothesis of molecular chaos is irrelevant. The H-theorem assumes that particle-particle collisions are extensive throughout the system. As depicted in Figure 1(d), before two particles randomly collide, their state is described by the two-body probability density ($\bar{f}_2$); after the collision, $\bar{f}_2$ is replaced by $\bar{f}_a \cdot \bar{f}_b$, where $\bar{f}_a$ and $\bar{f}_b$ are the one-body probability densities of the two particles, respectively. The information loss breaks time symmetry of the evolution of the probability of system state. Thus, although all the governing equations are time-reversible, entropy increase is irreversible. Yet, in a Knudsen gas, particle-particle collisions rarely happen.

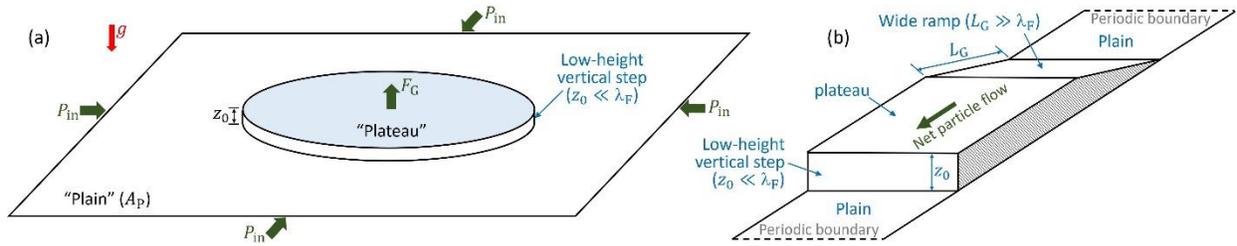

**Figure 2.** Two macroscopic non-thermodynamics models [13,15]. They are based on locally nonchaotic SND, unrelated to Feynman's ratchet or Maxwell's demon. **(a)** In a uniform gravitational field ($g$), a large number of elastic particles (not shown) randomly move on the upper shelf (the "plateau") and the lower shelf (the "plain") across the vertical step. The system is closed and immersed in a thermal reservoir. The plateau can be raised or lowered by the support force ($F_G$), and the plain ($A_P$) can be compressed or expanded by the in-plane pressure ($P_{in}$). The plateau height ($z_0$) is much less than the nominal particle mean free path ($\lambda_F$). In the step, as the particle-particle collisions are sparse, the particles tend to ascend or descend individually. A remarkable consequence is that at the steady state, the plateau-to-plain ratio of particle number density ($\bar{\rho}_G$) is inherently different from the Boltzmann factor ($\delta_0$). Because $\bar{\rho}_G \neq \delta_0$, by alternately operating $P_{in}$ and $F_G$ in an isothermal cycle, useful work can be produced by absorbing heat from the environment with no other effect. **(b)** A variant system. One end of the plateau is connected to the plain through the low-height step ($z_0 \ll \lambda_F$), and the other end is connected through the wide ramp. The ramp width $L_G \gg \lambda_F$. Across the chaotic ramp, the local $\bar{\rho}_G \to \delta_0$; across the nonchaotic step, the local $\bar{\rho}_G \neq \delta_0$. Thus, at the steady state, a net particle flow is spontaneously generated, without an energetic penalty. Notice that to achieve a nontrivial nonequilibrium effect, the gravitational field must be sufficiently strong. If the particles are ambient air molecules, $g$ needs to be at the level of neutron stars.

In Sections 2 and 3 below, case studies are performed on heat transfer in two Knudsen-gas models. In Section 2, a Knudsen gas is placed in gravity, similar to the locally nonchaotic energy barrier in [13]. In Section 3, by using switchable inner walls as time-dependent entropy barriers, a Knudsen-gas cell cluster is cyclically converted to a chaotic ideal gas. The numerical result of the second model is discussed in Section 4, and the first model is discussed in Section 5. Both models are compatible with the principle of maximum entropy, but break the boundaries of the second law of thermodynamics. Section 6 presents the considerations of experimental research.

Hereafter, the term "nonequilibrium" (or "intrinsically nonequilibrium") refers to a steady state that is spontaneously different from thermodynamic equilibrium; the system is closed and, unless otherwise specified (e.g., Section 2.2), immersed in a thermal bath. The out-of-equilibrium characteristics are rooted in the system's innate property of nonchaoticity (i.e., the lack of extensive particle-particle collisions), rather than being caused by fluctuations or any external driving force. The study is in the framework of classical mechanics, and mainly numerical. The computer programs are available at [16].

## 2. The first model: a Knudsen gas in gravity

In the past, people extensively studied heat transfer either in a nonchaotic medium without gravity [17-20] or in gravity for a chaotic medium [21-23]. For instance, when the system height ($D$) is much larger than the particle mean free path ($\lambda_F$), the planar Fourier flow associated with the chaotic particle movements agrees with the BGK model, and the average gas-phase temperature is equal to the average temperature of the environment. Loschmidt discussed the nonuniform temperature profile in a thin gas layer in gravity ($g$) [24]: when the height increases by $z_0$ ($z_0 < \lambda_F$), the average particle kinetic energy decreases by $mgz_0$, where $m$ is the particle mass. Loschmidt's setup does not exchange heat with the environment; it is cooler at the top and hotter at the bottom, contrary to our analysis in Equations (1,2) and Figure 3 below.

In this section, we investigate a model wherein not only the gas medium is nonchaotic ($D \ll \lambda_F$) but also the gravity effect is significant. The upper and lower boundaries are thermal walls. Such a setting is critically distinct from the conventional cases.

### 2.1 Thermally nonequilibrium steady state in a thermal bath

Figure 3(a) shows the Monte Carlo (MC) simulation of a two-dimensional (2D) gas, wherein a number of billiard-like particles randomly move in a uniform gravitational field ($g$) in a vertical plane. The 2D gas particles are finite-sized hard disks. They can collide with each other but there is no long-range force among them. The algorithm of the computer program is introduced in Section A2 in the Appendix.

The left and right borders (AA′ and BB′) are open and use periodic boundary condition. The upper and lower borders (AB and A′B′) are thermal walls. Each thermal wall represents the effects of a large thermal reservoir. The top-wall temperature and the bottom-wall temperature are denoted by $T_t$ and $T_b$, respectively. When a particle collides with a thermal wall, the reflected direction is random; the reflected particle speed is not correlated with the incident speed, but randomly follows the 2D Maxwell-Boltzmann distribution $p(v) = \beta m v \cdot e^{-\beta m v^2/2}$, where $\beta = 1/(k_B T_b)$ at the bottom boundary and $\beta = 1/(k_B T_t)$ at the top boundary. Such a thermal-wall boundary condition is commonly used in the study of rarefied gases [1,2]. Different boundary conditions are examined in Figure 4 and Section 4.3 below, as well as in [13]. The cause of the intrinsically nonequilibrium steady state is not any specific form of boundary condition, but rather the lack of extensive particle-particle collisions, i.e., nonchaoticity (also see Figure 5b,c).

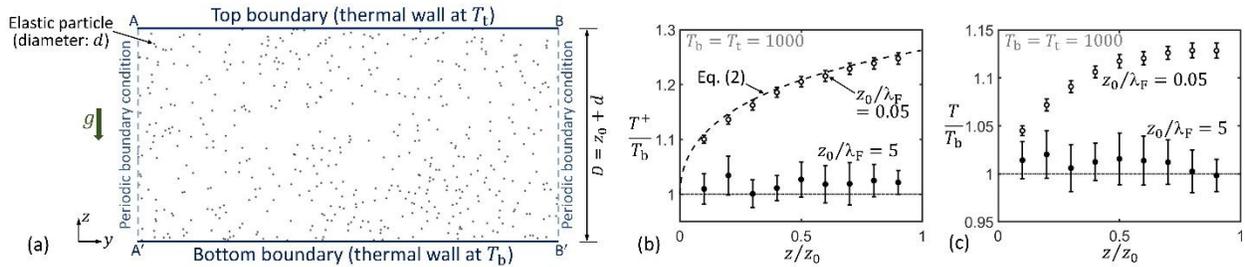

**Figure 3** The energy-barrier model: (**a**) a 2D gas in a vertical plane in a uniform gravitational field ($g$). When the effective plane height ($z_0 = D - d$) is less than the nominal particle mean free path ($\lambda_F$), the system is a Knudsen gas and cannot reach thermal equilibrium. (**b**) Typical steady-state distribution of the effective kinetic temperature ($T^+$) along height $z$, for the ascending particles. The bottom wall and the top wall are at the same temperature ($T_b = T_t = 1000$). The dashed curve is calculated from Equation (2). In the chaotic case ($z_0/\lambda_F = 5$), $T^+$ is uniform along $z$, i.e., the system is in thermal equilibrium. In the nonchaotic case ($z_0/\lambda_F = 0.05$), $T^+$ is nonuniform along $z$, i.e., thermal equilibrium cannot be reached. (**c**) Typical steady-state distribution of the effective kinetic temperature of all the particles ($T$). The trend of the $T - z$ relationship is similar to that of $T^+$. The error bars represent the 95%-confidence interval.

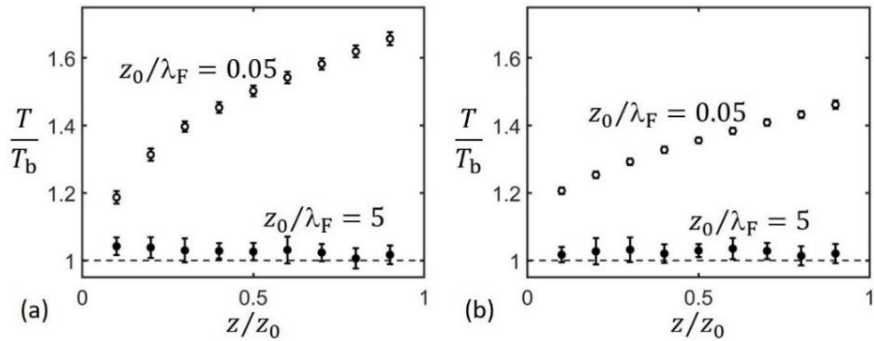

**Figure 4.** Effects of the boundary condition on the effective gas-phase kinetic temperature ($T$). (**a**) Both of the bottom and top boundaries are diffuse walls. The reflected particle direction is random; the reflected particle speed is equal to the incident speed. (**b**) The bottom boundary is a thermal wall (the same as in Figure 3); the top boundary is a diffuse wall (the same as in panel (a)). All the other settings and procedures are the same as in Figure 3. The error bars represent the 95%-confidence interval.

The bottom-wall temperature ($T_b$) is set to be the same as $T_t$. The gas phase can be viewed as being immersed in a thermal bath. The system is scalable; an example unit system is based on Å, fs, g/mol, and K. The particle diameter $d = 1$, which may be used as the normalization factor of length. The particle mass $m = 1$. The particle number $N = 500$. The timestep is 0.01. In all the simulation cases, $T_t = 1000$, and the total area of particle movement is $A_0 = w_0 z_0 = 39268.75$, where $z_0 = D - d$ is the effective plane height, and $w_0$ and $D$ are the width and the height of the simulation box, respectively. The nominal particle mean free path is $\lambda_F = A_0/(\sqrt{8}Nd) \approx 27.77$. Initially, the particles are randomly distributed in the plane. Their speed follows the 2D Maxwell-Boltzmann distribution at $T_b$, and their direction is random.

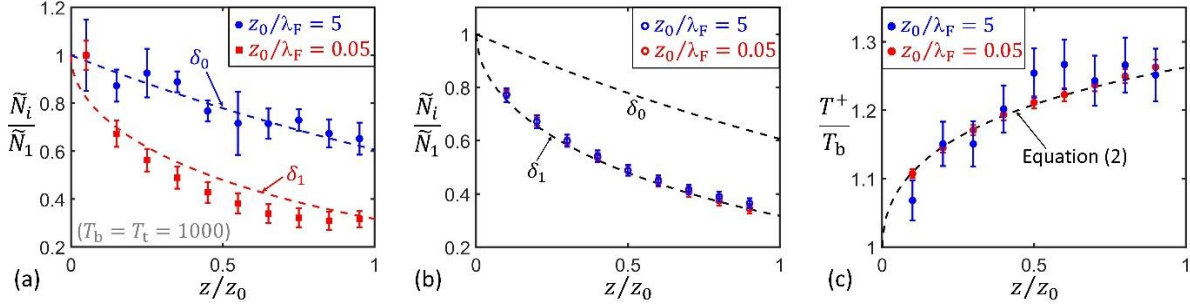

**Figure 5. (a)** The distribution of the particle flux ($\widetilde{N}_i$) along height $z$. The normalization factors ($\widetilde{N}_1$) are 56.10 and 103.73 for $z_0/\lambda_F = 5$ and $z_0/\lambda_F = 0.05$, respectively. **(b)** The reference simulation for "ghost" particles: the steady-state $\widetilde{N}_i$ and **(c)** the steady-state $T^+$ as functions of height $z$. The particle-particle collision is turned off. All the other parameters and procedures are the same as in Figure 3. The red and blue data points nearly overlap, indicating that particle-particle collision is the key factor. The error bars represent the 95%-confidence interval.

In different simulation cases, $z_0$ may be different, and $g$ is adjusted accordingly to maintain a constant Boltzmann factor at the top boundary $e^{-\beta m g z_0} = 0.607$. When $z_0 \gg \lambda_F$, the system represents a chaotic gas. When $z_0 \ll \lambda_F$, the system is a Knudsen gas and, because the particle-particle collisions are rare, its performance is dominated by the particle-wall collisions at the upper and lower borders.

After the settlement period ($10^5$ timesteps), we monitor the velocity of every particle for two cases: $z_0/\lambda_F = 0.05$ and $z_0/\lambda_F = 5$. Along the vertical direction ($z$), the plane is virtually divided into a number of equal horizontal layers. Figure 3(b) shows typical steady-state profiles of the effective kinetic temperature of the ascending particles: $T^+ = \overline{K}_i^+/k_B$, where $\overline{K}_i^+$ is the average kinetic energy of the particles moving upward across the lower border of the $i$-th horizontal layer (from bottom to top, $i = 1, 2, 3 \ldots$). If all the ascending and descending particles are considered, the overall effective temperature is given in Figure 3(c): $T = \overline{K}_i/k_B$, where $\overline{K}_i$ is the average kinetic energy of all the particles crossing the lower border of the $i$-th horizontal layer.

Maxwell investigated the gravity effect on thermal equilibrium of a column of chaotic gas and concluded [25]: "…the temperature would be the same throughout (i.e., isothermal), or, in other words, gravity produces no effect in making the bottom of the column hotter or colder than the top." This phenomenon is visualized in Figure 3(b,c): when $z_0/\lambda_F = 5$, both $T^+$ and $T$ tend to be homogeneous along $z$. It is consistent with Figure 5(a). The particle flux ($\widetilde{N}_i$) is defined as the number of the particles crossing the lower border of the $i$-th horizontal layer in every 1000 timesteps. If $z_0/\lambda_F = 5$, $\widetilde{N}_i$ fits well with the Boltzmann factor $\delta_0 = e^{-\beta mgz}$.

When $z_0/\lambda_F = 0.05$, remarkably, the steady-state distributions of $T^+$ and $T$ in Figure 3(b,c) are nonuniform, suggesting that without extensive particle-particle collisions, there is no mechanism to drive the system to reach thermal equilibrium. Correspondingly, the steady-state distribution of $\widetilde{N}_i$ is also non-Boltzmannian (Figure 5a). Because the horizontal-dimension particle kinetic energy has little contribution to the vertical particle movement, compared to the equilibrium case, fewer particles can overcome the gravitational energy barrier [13]. As a result, $\widetilde{N}_i$ is dominated by the $z$-component of particle velocity ($v_z$) and its distribution follows $\delta_1 = \int_{\sqrt{2gz}}^{\infty} p_z(v_z) dv_z = 1 - \text{erf}(\sqrt{\beta mgz})$ [13], where $p_z(v_z) = \sqrt{2\beta m/\pi}\, e^{-\beta m v_z^2/2}$ is the one-dimensional Maxwell-Boltzmann distribution of $|v_z|$.

It is counterintuitive that $T^+$ and $T$ increase with height $z$. On the one hand, the speed of every ascending or descending particle is lower at the top [24]. On the other hand, while the high-energy particles may reach the top boundary, many more low-energy particles can only move around the bottom boundary, i.e., the probability density distribution of particle kinetic energy varies with $z$. The two effects cannot counterbalance each other to render $T$ or $T^+$ independent of $z$; otherwise, thermal equilibrium (i.e., the second law of thermodynamics) would be equivalent to single-particle kinetics (e.g., Newton's second law), which contradicts the basic concept of thermodynamics such as the H-theorem. Overall, the height-dependent distribution of particle kinetic energy is more important, so that the average particle speed near the bottom is smaller than near the top. The average kinetic energy of the ascending particles can be calculated as

$$\bar{K}^+(z) = \bar{K}_y + \frac{\int_{mgz}^{\infty}(K_z - mgz)\cdot p_s(K_z) dK_z}{\int_{mgz}^{\infty} p_s(K_z) dK_z} \tag{1}$$

where $\bar{K}_y = k_B T_b/2$ is the average kinetic energy in the horizontal direction ($y$), and $p_s(K_z) = e^{-\beta K_z}/\sqrt{\pi K_z k_B T_b}$ is the one-dimensional Maxwell-Boltzmann distribution of the $z$-dimension kinetic energy ($K_z$) at $T_b$. Therefore,

$$T^+(z) \triangleq \frac{\bar{K}^+}{k_B} = \left[\frac{\Gamma(3/2, \beta mgz)}{\Gamma(1/2, \beta mgz)} - \frac{mgz}{k_B T_b} + \frac{1}{2}\right] T_b \tag{2}$$

where $\Gamma(x_1, x_2) \triangleq \int_{x_2}^{\infty} \xi^{x_1-1} e^{-\xi} \mathrm{d}\xi$ indicates the upper incomplete gamma function. Equation (2) reflects the intrinsic property of the Maxwell-Boltzmann distribution function, $p_\mathrm{s}(K_\mathrm{z})$.

In Figure 3(c), the gradient of $T$ is approximately a half of that of $T^+$ in Figure 3(b), which looks plausible, because $T$ accounts for the descending particles, while $T^+$ does not. We tested different boundary conditions in Figure (4). As long as $z_0/\lambda_\mathrm{F} \ll 1$, the $T-z$ gradient is always significant. Figure 5(b,c) shows the reference tests on "ghost" particles. When the particle-particle collision is turned off, with everything else remaining the same as in Figure 3, for both $z_0/\lambda_\mathrm{F} = 5$ and $z_0/\lambda_\mathrm{F} = 0.05$, the steady-state distributions of $\widetilde{N}_i$ and $T^+$ are nonequilibrium. It confirms that the dominant factor of the nonuniform $T$ is the lack of particle-particle collision.

2.2 Spontaneous cold-to-hot heat transfer

In Figure 6, $T_\mathrm{b}$ is varied, with everything else being the same as in Figure 3(b,c) ($z_0/\lambda_\mathrm{F} = 0.05$ and $T_\mathrm{t} = 1000$). The ratio of $T_\mathrm{b}/T_\mathrm{t}$ ranges from 0.6 to 1.0. At the steady state, the wall-to-gas heat transfer rate is calculated as $\phi = \Sigma_\mathrm{r}(K_\mathrm{re} - K_\mathrm{in})/\Delta t$, where $\Sigma_\mathrm{r}$ indicates summation for all the particles reflected by the upper wall or the lower wall in every $\Delta t = 2 \times 10^4$ timesteps, and $K_\mathrm{re}$ and $K_\mathrm{in}$ are the reflected particle kinetic energy and the incident particle kinetic energy, respectively. The reference heat transfer rate is $\phi_0 = \sqrt{2}(k_\mathrm{B}T_\mathrm{t})^{3/2}/(z_0\sqrt{m})$. Figure 7 shows typical time profiles of $\phi$ at the top and bottom boundaries.

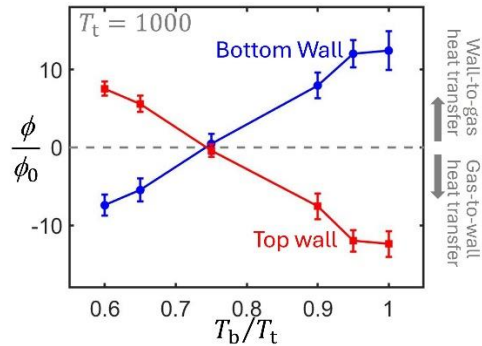

**Figure 6.** The steady-state wall-to-gas heat transfer rates ($\phi$) at the top and bottom boundaries, as functions of $T_\mathrm{b}$ ($T_\mathrm{t} = 1000$; $z_0/\lambda_\mathrm{F} = 0.05$). The error bars indicate the 95%-confidence interval. When $T_\mathrm{b}/T_\mathrm{t} = 0.9$ or 0.95, heat spontaneously transfers from the cold side (the bottom wall) to the hot side (the top wall) across the gas phase, without an energetic penalty. When $T_\mathrm{b}/T_\mathrm{t} = 0.75$, $\phi \approx 0$; i.e., effectively, the gas phase is thermally insulating.

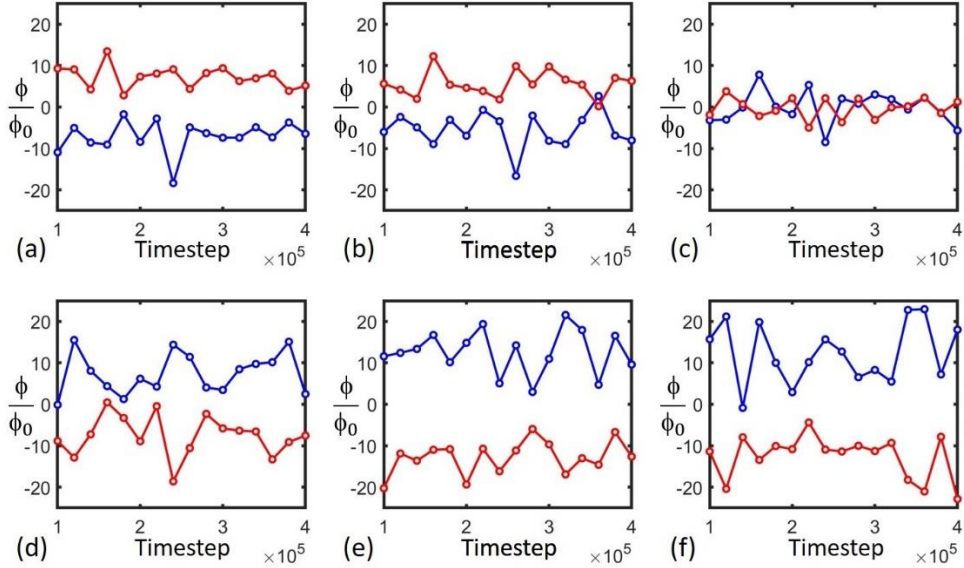

**Figure 7.** Typical time profiles of the wall-to-gas heat transfer rates at the top thermal wall (the red lines) and the bottom thermal wall (the blue lines): **(a)** $T_\text{b} = 600$; **(b)** $T_\text{b} = 650$; **(c)** $T_\text{b} = 750$; **(d)** $T_\text{b} = 900$; **(e)** $T_\text{b} = 950$; **(f)** $T_\text{b} = 1000$. In all the simulation cases, $T_\text{t} = 1000$ and $z_0/\lambda_\text{F} = 0.05$. In (d,e), heat transfers spontaneously from the cold side (the bottom wall) to the hot side (the top wall) across the gas phase. In (c), the heat flux is nearly zero.

When $T_\text{b}/T_\text{t} = 1$, the calculation reflects the nonequilibrium case in Figure 3(b,c). The heat influx at the top wall is dominated by $T^+$. As $T^+$ increases with $z$, the top wall absorbs heat from the incident particles. The "extra" thermal energy comes from the heat desorbed by the bottom wall. When $T_\text{b}/T_\text{t} = 0.9$ or $T_\text{b}/T_\text{t} = 0.95$, the positive $T^+ - z$ gradient overcomes the negative $T_\text{b} - T_\text{t}$ difference, causing a continuous heat transport from the cold side (the bottom wall at $T_\text{b}$) to the hot side (the top wall at $T_\text{t}$). When $T_\text{b}/T_\text{t} = 0.75$, the effects of the $T^+ - z$ gradient and the $T_\text{b} - T_\text{t}$ difference counterbalance each other. The overall heat flux is nearly zero and effectively, the gas phase is thermally insulating. Only when $T_\text{b}$ is significantly lower than $T_\text{t}$ ($T_\text{b}/T_\text{t} < 0.75$), can hot-to-cold heat transfer occur across the gas phase.

## 3. The second model: a switchable Knudsen-gas cell cluster

As discussed in the introductory section and detailed in Section A1 in the Appendix, at the steady state, the effective kinetic temperature of a Knudsen gas ($T$) is significantly lower than the container-wall temperature ($T_0$), i.e., a Knudsen gas cannot relax to thermal equilibrium. Since $T < T_0$, the internal energy of the Knudsen gas ($U$) is smaller than its equilibrium counterpart ($U_0$).

Notice that thermodynamic equilibrium is an accessible state. If a Knudsen gas is initially at the "equilibrium" state (at $T_0$), as it evolves to the nonequilibrium steady state (at $T$), in accordance with the first law of thermodynamics, the reduction in $U$ must be accompanied by heat desorption ($Q$). In this section, we perform a MC simulation to demonstrate such a process. In Section 4, we will show that the unusual thermal phenomenon is inconsistent with the second law of thermodynamics.

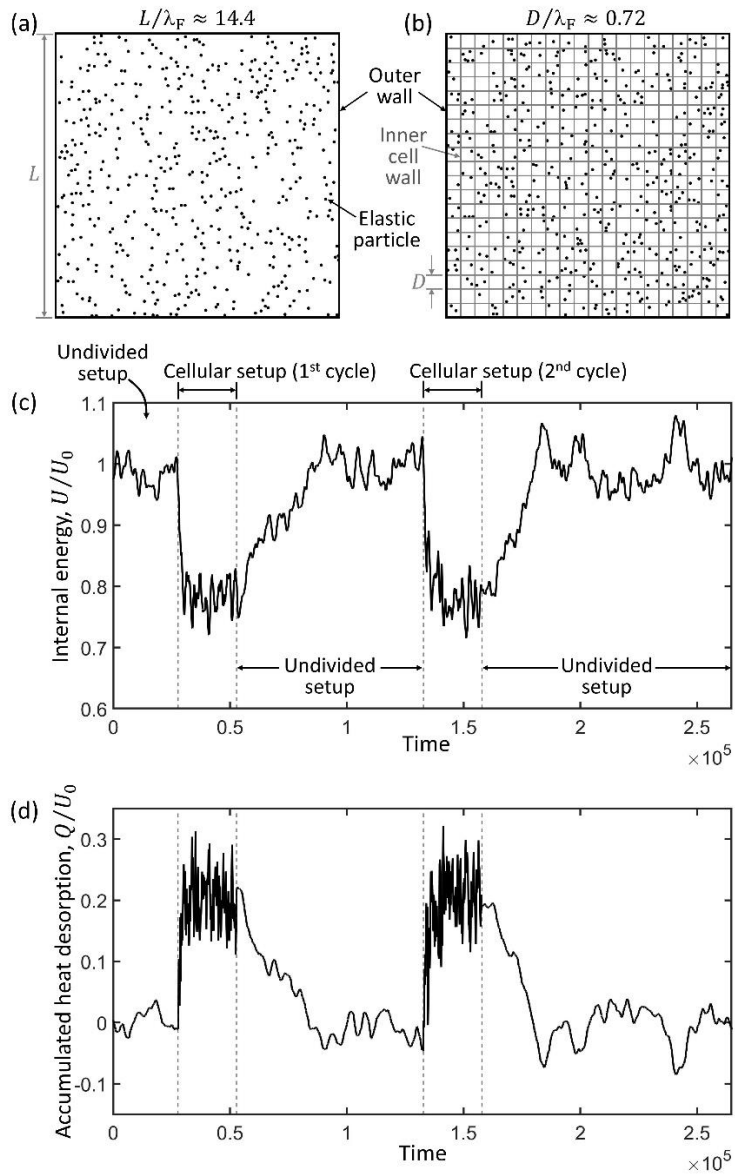

**Figure 8** The model of time-dependent entropy barriers: **(a)** the undivided setup (a 2D chaotic gas), and **(b)** the cellular setup (each cell is a Knudsen gas). Gravity is not considered. The outer and inner boundaries are thermal walls at a constant temperature ($T_0$). As the system shifts between the two configurations, the internal energy ($U$) changes, causing a significant heat flux through the walls. **(c)** A typical time profile of the internal energy ($U$) in two cycles of cell-wall insertion and removal ($U_0 = 0.029$). For each cycle, the first dashed vertical line indicates cell-wall insertion, and the second dashed vertical line indicates cell-wall removal. **(d)** The accumulated heat desorption across all the walls ($Q$), reported for every 100 particle-wall collisions. It matches the trend of $U$ in panel (c).

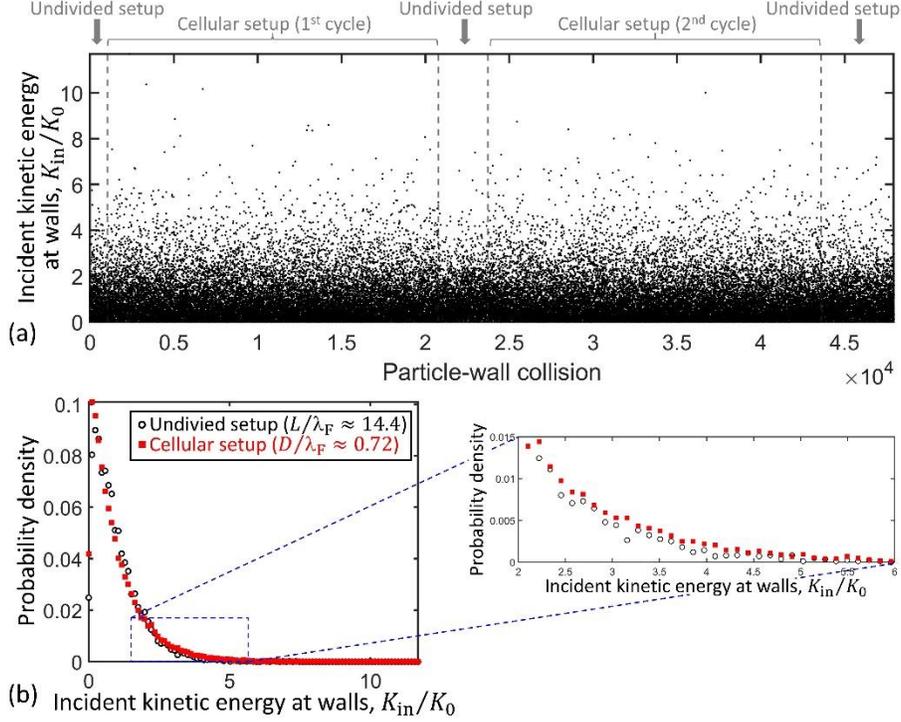

**Figure 9 (a)** The incident particle kinetic energy ($K_{in}$) at the container/cell walls, where $K_0 = k_B T_0$. Each data point represents a particle-wall collision. In the cellular setup, the particle-wall collisions happen more frequently than in the undivided setup. **(b)** The probability density distribution of $K_{in}$. The inset shows a magnified view for the high-energy range. It can be seen that in the cellular setup, compared to the undivided setup, the probability density of the high-energy particle-wall collisions ($K_{in}/K_0 \geq 2$) is larger.

Figure 8(a) depicts the simulation box of a 2D chaotic gas in a square container. Gravity is not considered. The unit system is the same as in Section 2. The container size ($L$) is 200; the particle number ($N$) is 500; the particle diameter ($d$) is 2; the particle mass ($m$) is 1; the timestep is 1. The mean free path of particle-particle collision is $\lambda_F = (L-d)^2/(\sqrt{8}Nd) \approx 13.9$, much less than $L$. The effective gas-phase kinetic temperature is defined as $T = \bar{K}/k_B$. The boundaries are thermal walls at constant $T_0 = 100$. When a particle collides with a wall, the reflected direction is random; the reflected particle speed is not correlated with the incident speed but instead randomly follows the 2D Maxwell-Boltzmann distribution of particle speed ($v$), $p_w(v) = \beta_0 m v \cdot e^{-\beta_0 m v^2/2}$, where $\beta_0 = 1/(k_B T_0)$. A variety of boundary conditions are tested in Section 4.3 below. The specific form of boundary condition does not affect whether the system can reach thermal equilibrium; the critical mechanism of $T < T_0$ is nonchaoticity (i.e., $D < \lambda_F$).

After the settlement period ($2 \times 10^4$ timesteps), at each timestep, internal energy ($U$) is calculated as the total kinetic energy of all the particles, and heat desorption ($Q$) is calculated as the accumulated $\Delta K$ of all the particle-wall collisions, where $\Delta K = K_{in} - K_{re}$. At the steady state, a set of divider walls are inserted, evenly separating the large container into 400 small cells (Figure 8b). The cell size ($D = 10$) is less than $\lambda_F$, so that each cell tends to be a Knudsen gas. The inner cell walls use the same boundary condition as the outer walls. The simulation of the particle movement continues. At the new steady state, the cell walls are removed, and the simulation keeps running. After the system returns to the initial steady state, a second cycle of cell-wall insertion and removal is operated.

Figure 8(c) shows the typical running average of $U$ for every 1000 timesteps. When the system is changed from Figure 8(a) to Figure 8(b), the steady-state $U$ decreases by about 20%, compatible with the literature data [9,10]. The reduction in $U$ ($\Delta U$) is significant but less than the ideal-case scenario of $T = T_0/2$ (Section A1 in the Appendix), which should be attributed to the occasional particle-particle collisions and the finite size of the simulation box. Corresponding to $\Delta U$, about $0.2 U_0$ heat is released from the gas phase into the thermal walls (Figure 8d). Figure 9 shows the probability density distribution of the kinetic energy of incident particles ($K_{in}$). In the cellular setup, compared to the undivided setup, high-energy particle-wall collisions happen more frequently, which explains how, at the steady state, although $T < T_0$, the gas-wall heat exchange is balanced. After the inner walls are removed, the system is converted back to the undivided state and absorbs $\sim 0.2 U_0$ heat from the environment. As the cell-wall operation is repeated, the heat desorption-absorption process continues cyclically.

**4. Discussion: time-dependent entropy barriers (the second model)**

4.1 Intrinsically nonequilibrium steady state

Figure 8(a) represents an ideal gas. It is a typical thermodynamic system and all its behaviors, including the inner cell walls in Figure 8(b), should be consistent with thermodynamic analysis. The cell walls block the particle movement and influence the probability of particle-particle collisions, and may be viewed as a type of entropy barrier. They are unrelated to Feynman's rachet [26] or Maxwell's demon [27]. As discussed in [14], Feynman's rachet is an equilibrium device, having no nonchaotic component; mere geometrical asymmetry does not offer any nonequilibrium mechanism. Maxwell's demon is subject to the energetic penalty of information processing [28]. In Figure 8(a,b), on the contrary, the operation of the cell walls can be pre-programed, without any knowledge of the specific position or momentum of each individual particle. Moreover, the heat flux in Figure 8(d) does not involve any random fluctuations or irreversible processes, e.g., the nonequilibrium paths associated with the Jarzynski equality [29].

In the cellular setup, heat desorption spontaneously occurs from the "cold" gas phase ($T < T_0$) to the "hot" walls (at $T_0$). No matter how to define temperature, the heat exchange can be substantial and drive a heat engine to produce useful work by absorbing thermal energy from the environment without any other effect, incompatible with the conventional heat-engine statement of the second law of thermodynamics.

4.2 Basic principle of maximum entropy

In this section, we show that the intrinsically nonequilibrium steady state, although counterintuitive, can be interpreted by the principle of maximum entropy [13,14]. It reassures us that the effective cold-to-hot heat transfer does not conflict with the fundamental logic: the most probable system state has the highest probability (measured by entropy).

For Figure 8(a,b), consider a canonical ensemble. Entropy is defined as

$$S = -k_B \sum_k \rho_k \ln \rho_k \qquad (3)$$

where $\rho_k$ is the probability of the $k$-th possible microstate. In Figure 8(a), the gas phase is chaotic and no detailed information is known about $\rho_k$, except for the following two constraints

$$\sum_k \rho_k = 1 \qquad (4)$$

$$\sum_k \rho_k \epsilon_k = U \qquad (5)$$

where $\epsilon_k$ is energy of the $k$-th possible microstate, and $U$ indicates the total particle kinetic energy at the steady state. Equations (3-5) define the Lagrangian [3]

$$\mathcal{L}_c = \left(-k_B \sum_k \rho_k \ln \rho_k + S_E\right) + \alpha_m \left(1 - \sum_k \rho_k\right) + \beta_m \left(U - \sum_k \rho_k \epsilon_k\right) \qquad (6)$$

with $\alpha_m$ and $\beta_m$ being the Lagrange multipliers. The constrained maximation of entropy ($S$) can be expressed as

$$\frac{\partial \mathcal{L}_c}{\partial \rho_k} = 0 \tag{7}$$

The solution of Equation (7) is $\rho_k = e^{-(k_B + \alpha_m + \beta_m \epsilon_k)/k_B}$. According to Equation (4), $\rho_k$ may be written as

$$\rho_k = \frac{1}{Z_e} e^{-\beta_n \epsilon_k} \tag{8}$$

where $Z_e = \sum_k e^{-\beta_n \epsilon_k}$ is the partition function, and $\beta_n = \beta_m / k_B$. Substitution of Equations (4,5,8) into Equation (3) gives the maximum possible entropy that the system can ever reach, i.e., the global maximum in the phase space that corresponds to thermodynamic equilibrium,

$$S_{eq} = k_B \ln Z_e + k_B \beta_n U \tag{9}$$

Since $\frac{\partial S_{eq}}{\partial U} = \frac{1}{T_0}$, $\beta_n = \beta_0$. Any other $\rho_k$ (e.g., Equation 13 below) would result in a smaller entropy than $S_{eq}$, corresponding to a non-Boltzmann state.

When the gas phase is formed by the Knudsen-gas cells (Figure 8b), the nonchaotic particle movements are less random than in the chaotic setup; that is, more information of $\rho_k$ is available. To derive analytical solutions, in this section, for the sake of simplicity, particle-particle collision is ignored. In the $k$-th possible microstate, for the $j$-th gas particle, denote its traveling distance and speed by $\lambda_{kj}$ and $v_{kj}$, respectively; $t_{kj} = \lambda_{kj}/v_{kj}$ is the time duration between the previous and the next particle-wall collisions. Similarly to the concept of $\rho_L \propto 1/v$ in the single-particle case in the introductory section, besides Equations (4) and (5), $\rho_k$ obeys additional rules: $\rho_k \propto t_{kj}$ (for all $j = 1, 2, \dots N$), which may be formulated as

$$\rho_k = \xi_0 \prod_{j=1}^N \xi_{kj} t_{kj} = C_k \tau_k \tag{10}$$

where $\xi_0$ is the normalization factor, $\xi_{kj}$ is the coefficient of $t_{kj}$, $C_k = \xi_0 \prod_{j=1}^N \xi_{kj}$, and $\tau_k = \prod_{j=1}^N t_{kj}$. Equations (3-5) and (10) re-define the Lagrangian

$$\mathcal{L}_n = \mathcal{L}_c + \sum_k b_k (\rho_k - C_k \tau_k) \tag{11}$$

where $b_k$ are the additional Lagrange multipliers. To maximize $S$, through

$$\frac{\partial \mathcal{L}_n}{\partial \rho_k} = 0 \tag{12}$$

we have

$$\rho_k = \varphi_m e^{\zeta_k} e^{-\beta_n \epsilon_k} \tag{13}$$

where $\varphi_m$ is the normalization factor, and $\zeta_k = \frac{b_k}{k_B}(1 - \frac{\partial C_k}{\partial \rho_k}\tau_k)$. Comparison between Equations (13) and (10) suggests that $C_k \propto e^{-\beta_n \epsilon_k}$ and $e^{\zeta_k} \propto \tau_k$. Consequently,

$$\rho_k = \frac{1}{Z_n} \tau_k e^{-\beta_n \epsilon_k} \quad (14)$$

where $Z_n = \sum_k (\tau_k e^{-\beta_n \epsilon_k}) \propto 1 - k_B \zeta_k / b_k$ is the generalized partition function. The form of $\rho_k \propto \tau_k e^{-\beta_n \epsilon_k}$ is in line with $\rho_L \propto p_w/v$. As the reflected particle speed at the thermal walls follows the Maxwell-Boltzmann distribution at $T_0$, it is reasonable to set $\beta_n$ equal to $\beta_0$. Based on Equations (14) and (5),

$$U = -\frac{\partial \ln Z_n}{\partial \beta_n} \quad (15)$$

In Equation (14), at a given energy level $\epsilon_k$, $\rho_k$ is not constant but rather varies with $\tau_k$. It is non-Boltzmannian and does not satisfy Boltzmann's assumption of equal *a priori* equilibrium probabilities.

On the one hand, as the derivation of Equation (14) is based on Equation (12), entropy is maximized. Combination of Equations (3), (14), and (15) gives the nonequilibrium entropy

$$S_{ne} = k_B \ln Z_n + k_B \beta_n U - k_B \bar{B}_k \quad (16)$$

where $\bar{B}_k = \sum_k (\rho_k \sum_{j=1}^N \ln t_{kj})$; for indistinguishable particles, $\bar{B}_k = N \sum_k (\rho_k \ln t_{kj})$ (for any $j$). On the other hand, the maximum possible entropy (i.e., the global maximum in the phase space) is associated with the equilibrium $\rho_k$ in Equation (8), not the nonequilibrium $\rho_k$ in Equation (14). Thus, $S_{ne} < S_{eq}$. The root cause of $S_{ne} < S_{eq}$ is that, compared to Equation (9), Equation (16) involves more restrictions on the maximization of entropy (Equation 10), so that $S_{ne}$ is the local maximum in the subregion of Equation (10) in the phase space.

To account for the intrinsically nonequilibrium steady state and also to remain consistent with the principle of maximum entropy, we may generalize the second law of thermodynamics as follows [14]: in an isolated system, the difference between entropy ($S$) and the maximum possible steady-state entropy ($S_Q$) cannot increase, i.e.,

$$S \to S_Q \quad (17)$$

For a chaotic system, $S_Q = S_{eq}$, so that $S \to S_Q$ is equivalent to the conventional statement of the second law of thermodynamics. For an intrinsically nonequilibrium system, $S_Q = S_{ne}$, which is less than $S_{eq}$. If initially $S > S_Q$, without an external driving force, entropy could spontaneously decrease. Equation (17) is the fundamental mechanism behind the non-thermodynamic thermal phenomena in Sections 2 and 3.

It is worth noting that if the thermal bath is finite-sized, when the undivided setup (Figure 8a) is shifted to the cellular setup (Figure 8b), the heat transfer from the gas phase to the thermal bath would cause the thermal-bath temperature ($T_0$) to rise, accompanied by an entropy increase ($\Delta S_c$). However, $\Delta S_c$ occurs after the triggering event of the nonequilibrium state (the cell-wall insertion). It does not affect the overall entropy variation between the two steady states. As long as thermal equilibrium cannot be reached ($T \neq T_0$), Figure 8(b) must have a smaller entropy than Figure 8(a), because of the higher degree of nonuniformity. Hence, as the system switches from the undivided setup to the cellular setup, the overall entropy decreases, without incurring an energetic penalty. Regardless of the characteristics of entropy, the heat flux ($Q$) can drive a heat engine to produce useful work in a cycle by absorbing heat from the thermal bath.

4.3 The cell-wall boundary condition

The thermally nonequilibrium steady state ($T < T_0$) does not rely on any specific characteristics of the boundary condition used in the computer simulation, particularly the probability density of the reflected particle speed at the container/cell walls ($p_w$). In a nonchaotic Knudsen-gas cell, as the particle-particle collisions are sparse, the effective gas-phase kinetic temperature ($T$) tends to be proportional to $\int_0^\infty (mv^2/2)(p_w/v)dv$. No matter what the specific function of $p_w(v)$ is, $T$ must be smaller than the cell-wall temperature $T_0 = k_B^{-1} \int_0^\infty (mv^2/2)p_w dv$, because the factor of $1/v$ in the integrand of $T$ gives more weight to the slower particles. The only function that can keep $T = T_0$ is $p_w(v) = 0$, which is trivial.

In Section 3, $p_w(v)$ is taken as the 2D Maxwell-Boltzmann distribution. In a numerical experiment, different forms of $p_w(v)$ are tested. When a particle collides with a wall, the reflected direction is random, and the reflected particle speed ($v_{re}$) is determined through $v_{re}^2 = \alpha_{in} v_{in}^2 + (1 - \alpha_{in})v_{rd}^2$, where $v_{in}$ is the incident speed, $v_{rd}$ is a speed randomly generated from the 2D Maxwell-Boltzmann distribution at $T_0$, and $\alpha_{in}$ represents the "memory" of the particle-wall interaction. In different simulation cases, $\alpha_{in}$ is varied from 0 to 1/2. When $\alpha_{in} = 0$, $v_{re}$ is reduced to the ideal thermal-wall condition in Section 3; when $\alpha_{in} > 0$, the degree of randomness of the particle-wall collision is lower. The simulation results suggest that in the range of $\alpha_{in}$ under investigation, $\alpha_{in}$ has no statistically significant influence on the steady state. It does affect the rate of convergence. With a larger $\alpha_{in}$, the system tends to take more time to reach the steady state, and vice versa.

In another numerical experiment on the cellular setup, $\alpha_{in}$ is set to 1 (i.e., $v_{re} = v_{in}$). The reflected angle is either random (diffuse reflection) or equal to the angle of incidence (specular reflection). In both cases, the system is isolated, and the internal energy does not vary. The average kinetic energy of the incident particles at the walls is significantly higher than the average kinetic energy of all the particles in the gas phase.

In yet another numerical experiment on the undivided setup, particle-particle collision is turned off. The particle-wall collision is not affected ($\alpha_{in} = 0$). At the container walls, significant heat desorption is observed, similarly to the heat exchange at the cell walls in the cellular setup. It confirms that the lack of extensive particle-particle collisions is the key factor of the spontaneously nonequilibrium system behavior.

The thermal-wall boundary condition is compatible with the basic concept of path independence of thermodynamics. It represents the effects of a chaotic body (the thermal bath) and has no nonequilibrium mechanism, i.e., it cannot be responsible for the nonequilibrium phenomena. To further understand the ideal case of $\alpha_{in} = 0$, we may assume that the cell walls are equipped with a layer of perfect heat exchangers. During a particle-wall interaction event, the particle fully exchanges kinetic energy with the thermal bath through the heat exchangers, before it departs back into the gas phase.

**5. Discussion: nonchaotic energy barrier (the first model)**

5.1 Intrinsically nonequilibrium steady-state temperature distribution

According to Figures 3 and 6, the nonchaotic system cannot reach thermal equilibrium. At the steady state, when $T_b = T_t$ (i.e., the Knudsen gas is immersed in a thermal bath), not only is the temperature distribution nonuniform, but also the heat exchange at the upper and lower boundaries ($Q$) is significant. When $T_b/T_t = 0.75$, $Q \approx 0$, i.e., the gas phase is effectively thermally insulating, despite the large difference between $T_b$ and $T_t$. This phenomenon is consistent with Equation (2) and Figure 3(b) that $T_b/T^+$ is close to 0.75 at the upper boundary ($z = z_0$). When $0.75 < T_b/T_t < 1$, the cold-to-hot heat transfer is spontaneous, incompatible with the refrigeration statement of the second law of thermodynamics. Moreover, as depicted in Figure 10, because the $T - z$ gradient is dependent on the gravitational energy barrier ($mgz_0$), two different gases can form Maxwell's double-column engine (Section A3 in the Appendix [30]), which continuously produces useful work by absorbing heat from the environment. It conflicts with the heat-engine statement of the second law of thermodynamics.

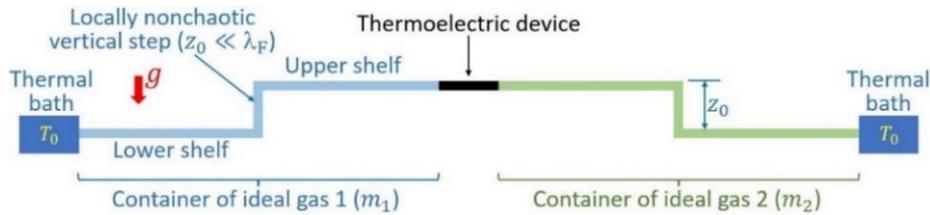

**Figure 10.** The two-shelf model inspired by Maxwell's double-column engine [30]. Two identical containers contain different ideal gases, in a uniform gravitational field ($g$). The gas-particle masses in the two containers are $m_1$ and $m_2$, respectively ($m_1 \neq m_2$). Each container consists of an upper shelf and a lower shelf, with the height difference ($z_0$) much less than the nominal particle mean free path ($\lambda_F$). The lower shelves of the two containers are in contact with the same thermal bath at $T_0$. A thermoelectric device is sandwiched in between the two upper shelves. As the temperature field is nonuniform along the vertical direction, because $m_1 \neq m_2$, the effective gas-phase temperatures at the two upper shelves are different. Thus, the thermoelectric device can continuously produce useful work by absorbing heat from the thermal bath without any other effect.

There are a few points worth noting. First, in Figure 6, heat desorption of the gas particles at the upper boundary is balanced by heat absorption at the lower boundary, obeying the first law of thermodynamics. Secondly, the system does not consume energy from the gravitational field. At the steady state, on average, for every ascending particle, there is a descending particle; vice versa. Thirdly, Figure 3(a) is fundamentally different from Earth's atmosphere. In [31], Maxwell stated: "In the case of the atmosphere, the effect of wind is to cause the temperature to vary as that of a mass of air would do if it were carried vertically upwards, expanding and cooling as it ascends." Earth's atmosphere is in contact with both the ground and outer space, leading to macroscopic air convection. As air rises, it undergoes adiabatic expansion and tends to be colder. On the contrary, in Figure 3(a), when $T_b = T_t$, the system is immersed in a thermal bath, and there is no planar Fourier flow. Moreover, without extensive particle-particle collisions, a volume increase would not do work, like free expansion in vacuum. That is, a Knudsen gas has no mechanism to render the upper section cooler. In fact, in Figure 3(b,c), $T^+$ and $T$ are higher at the top.

5.2 Entropy associated with the intrinsically nonequilibrium steady state

In Figure 3, for the $k$-th possible microstate, use $v_{yj}$, $v_{zj}$, and $z_j$ to denote the horizontal component of velocity, the vertical component of velocity, and the height of the $j$-th gas particle, respectively ($j = 1,2 \ldots N$). The nonchaotic particle movement imposes additional constraints on $\rho_k$. Following the concept of Equation (14) and $\rho_L \propto p_w/v$ (Section A1 in the Appendix), if particle-particle collision is ignored, $\rho_k$ may be written as

$$\rho_k = \varphi_t \Phi_k \bar{\Phi}_k \qquad (18)$$

where $\varphi_t$, $\Phi_k$, and $\bar{\Phi}_k$ are defined for the $k$-th possible microstate of the gas phase; $\Phi_k = \prod_{j=1}^{N_b} p_b(v_{yj}) \cdot p_b(v_{1j}) \cdot t_{aj}$ represents the contribution of the particles reflected into the vertical plane from the bottom boundary; $\bar{\Phi}_k = \prod_{j=1}^{N_t} p_t(v_{yj}) \cdot p_t(v_{2j}) \cdot t_{dj}$ represents the contribution of the particles from the top boundary; $N_t$ and $N_b$ are the numbers of the particles from the bottom boundary and the top boundary, respectively ($N_t + N_b = N$); $\varphi_t = \varphi_a \varphi_k$; $\varphi_k$ is the probability of $N_t$; $\varphi_a = 1/(\sum_k \varphi_k \Phi_k \bar{\Phi}_k)$ is the normalization factor; $p_b$ and $p_t$ denote the one-dimensional Maxwell-Boltzmann distributions of speed at the bottom boundary (at $T_b$) and the top boundary (at $T_t$), respectively; $v_{1j}$ and $v_{2j}$ are respectively the initial $v_{zj}$ after the previous particle-wall collisions at the bottom boundary and the top boundary, determined through $v_{1j}^2 = v_{zj}^2 + 2gz_j$ and $v_{2j}^2 = v_{zj}^2 - 2g(z_0 - z_j)$; $t_{aj}$ and $t_{dj}$ denote the time durations between the previous and the next particle-wall collisions of the $j$-th particle. For the particles from the bottom boundary, if $v_{1j} < \sqrt{2gz_0}$, $t_{aj} = 2v_{1j}/g$; if $v_{1j} \geq \sqrt{2gz_0}$, $t_{aj}$ is determined through $(v_{1j} - t_{aj}g)^2 = v_{1j}^2 - 2gz_0$. For the particles from the top boundary, $t_{dj}$ is determined through $(v_{2j} + t_{dj}g)^2 = v_{2j}^2 + 2gz_0$.

Based on Equation (3), Equation (18) gives the nonequilibrium entropy

$$S_{ne} = -k_B \sum_k [\varphi_t \Phi_k \bar{\Phi}_k \cdot \ln(\varphi_t \Phi_k \bar{\Phi}_k)] \qquad (19)$$

Similar to the discussion in Section 4.2, because of the nonchaoticity constraints on $\rho_k$ associated with $t_{aj}$ and $t_{dj}$, $S_{ne}$ is a local maximum in the phase space, less than its chaotic counterpart.

## 6. Considerations about experimental research

The non-thermodynamic phenomena in Sections 2 have never been reported for real-world systems, which should be attributed to the weak gravitational field near Earth ($\beta mgz_0 \approx 10^{-12}$ for ambient air molecules). It would be interesting to explore whether the concept can be nontrivial in a high-$g$ environment (e.g., near a neutron star or a blackhole), at the astronomical scale, by using a powerful centrifuge (e.g., a heavy gas at a low temperature), or with a stronger thermodynamic force (e.g., the Coulomb force). Notice that if a conductive or semiconductive nanolayer is placed in an electric field, the screening effect must be taken into consideration, e.g., by keeping the characteristic size smaller than the Debye length or the spacing of the charge carriers. Cooper pairs could have unique properties, and their role in heat transfer is worth investigating. Other operation mechanisms [e.g., 13-15] and weakly/sparsely interacting particles are also important research topics.

Section 3 circumvents the hurdle of the large energy barrier. However, the idealized process of cell-wall insertion and removal is difficult to achieve, because gas transport is influenced by surface adsorption [32]. This issue may be resolved by using two identical chambers. One set of divider walls are alternately inserted and removed between the two chambers. Thus, the adsorbed gas particles only have internal effects.

Interestingly, in a study on nonwetting liquids in nanopores [33,34], there may have already been experimental evidence of the entropy-barrier effect. Because the nanopore walls are not wettable to the liquid, the surface adsorption problem is minimized. In [34] (see Figure 11a), 1 g hydrophobic nanoporous silica particles were immersed in 5 g aqueous solution of sodium chloride, sealed in a steel container with a thermal insulation layer. A pressure ($P$) was applied on the liquid phase through the piston, so that the liquid was forced to fill the nanopores. Then, $P$ was gradually reduced, accompanied by liquid defiltration. The temperature of the bulk liquid phase was monitored by the embedded sensor. Thermodynamic analysis suggests that separation of a nonwetting liquid from a solid surface has a tendency to cause a temperature increase, rather than a temperature decrease. Yet, a large reduction in temperature ($\Delta T_s \approx -1.5$ ºC) was measured (Figure 11b). It corresponds to a loss of thermal energy $U_L \approx 30$ J, while the total work done by the piston ($E_P$) is only about 16 J. Even if $E_P$ could cause cooling and the efficiency were 100%, $|\Delta T_s|$ should not exceed 0.7 ºC. This phenomenon may be attributed to the "Knudsen gas like" characteristics of the confined liquid. As the pore size is comparable with the mean free path of the liquid molecules, the effective kinetic temperature of the confined liquid is less than the nanopore-wall temperature. Thus, when the confined liquid moves out of the nanopores, the overall temperature becomes lower. Likewise, liquid infiltration resulted in a large temperature increase $\Delta T_s \approx 2.1$ °C [34]; the associated gain of thermal energy ($U_A \approx 44$ J) was much greater than $E_P$. The difference between $U_A$ and $U_L$ was close to $E_P$, consistent with the fact that $E_P$ was mostly dissipated through hysteresis. Figure 11(c) illustrates a design of circular flow, wherein the steady-state liquid-solid interaction does not vary over time. The resistance to the movement of the nonwetting liquid may be low, thanks to the superfluidity effect in nanoenvironment [35-37].

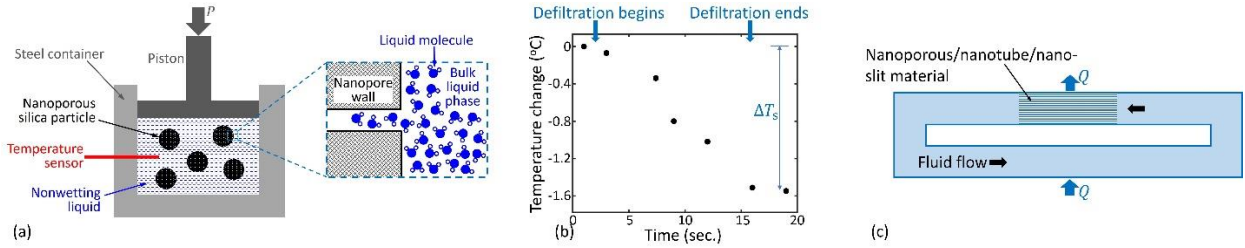

**Figure 11. (a)** The experimental setup in [33,34]. The temperature variation associated with liquid defiltration from the nanoporous silica particles is measured. **(b)** Typical testing data [34]: temperature considerably decreases when the nonwetting liquid defiltrates out of the nanoporous silica. **(c)** Schematic of a nanofluidic setup of continuous flow, based on the same concept as in panel (a). The system is closed and immersed in a thermal bath. The effective nanochannel size is smaller than or comparable with the mean free path of the fluid particles. As the fluid enters the nanochannels, it releases heat ($Q$) to the environment, because of the nonequilibrium effect. Likewise, when the fluid comes out of the nanochannels, it absorbs heat from the environment.

To apply the concept of entropy barriers to the charge carriers in a mesoscopic physical system, "cell walls" may be formed by using adjustable potential wells, outer/inner boundaries, asymmetric scattering or reflection, etc. The configuration in Figure 11(c) can be tested by using conductive nanowires, nanostrips, nanotubes, and/or nanolayers. The particle-barrier interaction must involve heat exchange (e.g., the barrier should not behave as a specular wall). The upper limit of the power density could be more than 10 kW/cm$^3$ (Section A4 in the Appendix). Multiple units may be connected in parallel or in tandem, to amplify the temperature difference and the heat flux.

## 7. Concluding remarks

It has long been known that certain nonchaotic particle movements cannot be analyzed by thermodynamics; usually, their energy properties are considered "trivial." In current research, we demonstrate that beyond the boundaries of Boltzmann's H-theorem, there are nontrivial non-thermodynamic systems with unusual heat transfer properties. One example is the Knudsen gas in gravity (Figure 3), and the other example is the switchable Knudsen-gas cell cluster (Figure 8). In both cases, with no other effect, heat can spontaneously transfer from the cold side to the hot side, allowing for production of useful work by absorbing heat from a single thermal reservoir. Such counterintuitive phenomena are incompatible with the conventional refrigeration statement and heat-engine statement of the second law of thermodynamics.

A Knudsen gas is defined as a rarified gas with the particle mean free path larger than the characteristic size of the gas container. Without extensive particle-particle collisions, the particle trajectories tend to be independent of each other, and the system state is dominated by the particle-boundary collisions. When a Knudsen gas is immersed in a thermal bath, it cannot relax to thermal equilibrium. In the first model of the Knudsen gas in gravity, the effective gas-phase kinetic temperature ($T$) is spontaneously nonuniform along height. In the second model of the Knudsen-gas cell cluster, $T$ remains significantly lower than the thermal-bath temperature. The intrinsically nonequilibrium steady states are not caused by fluctuations or any external driving force, but rather rooted in the innate property of nonchaoticity. As the hypothesis of molecular chaos is inapplicable, entropy is still maximized, but it reaches a local maximum in the phase space, smaller than the global maximum (the equilibrium entropy).

In the future, the detailed transition mechanism between nonchaotic and chaotic behaviors needs to be further explored. Other interesting topics of study include the generalized rule of thermodynamic analysis, implications for quantum mechanical systems, experimental design and verification, to name a few.

**Appendix**

A1. Knudsen gas: intrinsically nonequilibrium steady state

The numerical simulations in Sections 2 and 3 take into account the particle-particle collisions. The particles collide with each other whenever they meet. However, in a Knudsen gas, compared to the particle-wall collisions, because the particle-particle collisions are sparse, their effect is secondary. In this section, for the sake of simplicity, to obtain analytical solutions, particle-particle collision is ignored.

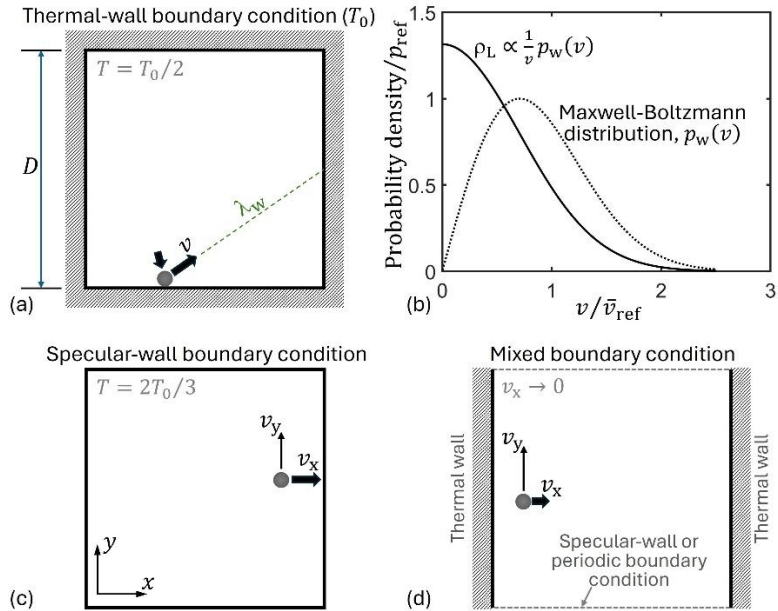

**Figure 12.** Three 2D Knudsen-gas models that exhibit thermally nonequilibrium steady states. **(a)** A Knudsen-gas cell, with the cell boundaries being made of thermal walls. The wall temperature is constant $T_0$; $\lambda_w$ indicates the average traveling distance between two consecutive particle-wall collisions. At the steady state, the effective gas-phase temperature $T = T_0/2$. **(b)** Comparison between $p_w(v)$ and $\rho_L \propto p_w(v)/v$, as functions of the particle speed ($v$). The normalization factor ($p_{ref}$) is the peak value of $p_w$, and $\bar{v}_{ref} = \sqrt{2k_B T_0/m}$. **(c)** A Knudsen-gas cell, with the boundaries being made of specular walls. At the steady state, $T = 2T_0/3$. **(d)** A Knudsen-gas cell. The two cell borders normal to the x-axis are thermal walls; the borders normal to the y-axis are specular walls or use periodic boundary condition. At the steady state, $v_x \to 0$.

Such a simplification may be realized in Figure 8(b): in the cellular setup, if the number of the cells is much larger than the total number of the particles, the system state would be dominated by the cells that contain only one particle each. The few crowded cells remain chaotic and do not contribute to the heat exchange. The empty cells are also trivial. In a single-particle cell, it is impossible for the particle to encounter another particle, and the effective mean free path of particle-particle collision $\lambda_F \to \infty$. The mean free path of particle-wall collision is $\lambda_w = \pi D/4$ [12], with $D$ being the cell size. Hence, $\lambda_F \gg \lambda_w$.

Figure 12(a) depicts a two-dimensional (2D) Knudsen-gas cell. The cell boundaries are thermal walls at a constant temperature ($T_0$). The reflected particle speed ($v$) follows the 2D Maxwell-Boltzmann distribution function $p_w(v) = \beta_0 m v \cdot e^{-\beta_0 m v^2/2}$, where $\beta_0 = 1/(k_B T_0)$. The reflected direction is random. Use $\rho_L(v)$ to denote the probability density of finding a particle at a speed $v$. Without particle-particle collision, $\rho_L$ is proportional to the retention time $\lambda_w/v$, i.e., the average time duration between two consecutive particle-wall collisions. Therefore, $\rho_L(v) \propto p_w(v)/v$, and its normalized form is

$$\rho_L = \sqrt{\frac{2\beta_0 m}{\pi}} e^{-\beta_0 m v^2/2} \tag{20}$$

Figure 12(b) shows $\rho_L(v)$ and $p_w(v)$. The average particle speed of $\rho_L$ is lower than that of $p_w(v)$, as it should be. Compared to the equilibrium state, it is more likely to find a slow particle in the cell, since the time interval between particle-wall collisions is inversely proportional to the particle speed ($v$). No reasonable boundary condition can keep the system at thermal equilibrium, i.e., the time-average particle kinetic energy ($\overline{K}$) cannot be equal to $k_B T_0$. The effective gas-phase kinetic temperature is

$$T = \frac{1}{k_B} \int_0^\infty \frac{m v^2}{2} \rho_L dv = \frac{1}{k_B} \int_0^\infty \left[\frac{m v^2}{2}\left(\sqrt{\frac{2\beta_0 m}{\pi}} e^{-\beta_0 m v^2/2}\right)\right] dv = \frac{T_0}{2} \tag{21}$$

In the numerical simulation in Section 3, $T$ is significantly less than $T_0$, but because of the occasional particle-particle collisions and the finite size of the simulation box, it is larger than the ideal-case scenario of $T_0/2$.

In comparison, in a chaotic gas (e.g., a large body of rarefied gas), due to the random particle-particle collisions, $\rho_L \propto p_w(v)/v$ is inapplicable. No particle can reach the boundary without interacting with other particles. In a short time period $t_n$, the particle influx at the boundary (i.e., the number of particle-wall collisions per unit boundary length) can be calculated as $\overline{N}_n = \rho_0(\bar{v}_x t_n)/2$, where $\rho_0$ is the particle number density, $\bar{v}_x$ is the average $v_x$, and $v_x$ is the component of incident particle velocity normal to the boundary. The average kinetic energy of the incident particles is

$$\overline{K}_{\text{in}} = \frac{1}{\overline{N}_{\text{n}}}\left[\overline{N}_{\text{n}} \int_0^\infty \frac{mv^2}{2}(\beta_{\text{T}} mv \cdot e^{-\beta_{\text{T}} mv^2/2}) dv\right] = k_{\text{B}} T \tag{22}$$

where $\beta_{\text{T}} = 1/(k_{\text{B}} T)$. Consequently, at the steady state, $T_0 = \overline{K}_{\text{in}}/k_{\text{B}} = T$. That is, thermal equilibrium is reached.

Figure 12(c) shows a 2D Knudsen-gas cell formed by specular walls. The particle movements in the x- and y-axes are independent of each other. Consider an initial condition that $v$ follows the Maxwell-Boltzmann distribution. The effective particle influx at the boundary is not $\overline{N}_{\text{n}}$, but rather $\overline{N}_{\text{c}} = (1/2) \int_0^\infty \rho_0 v_{\text{x}} t_{\text{n}} p_{\text{v}}(v_{\text{x}}) dv_{\text{x}}$, where $p_{\text{v}}(v_{\text{x}}) = \sqrt{2\beta_{\text{T}} m/\pi}\, e^{-\beta_{\text{T}} mv_{\text{x}}^2/2}$ is the one-dimensional Maxwell-Boltzmann distribution function of $|v_{\text{x}}|$. At the cell wall, the average kinetic energy of the incident particles is

$$\overline{K}_{\text{in}} = \frac{1}{\overline{N}_{\text{c}}}\left\{\frac{1}{2}\int_0^\infty \frac{mv_{\text{x}}^2}{2}[\rho_0 v_{\text{x}} t_{\text{n}} p_{\text{v}}(v_{\text{x}})] dv_{\text{x}}\right\} + \overline{K}_{\text{pr}} = \frac{3}{2} k_{\text{B}} T \tag{23}$$

where $\overline{K}_{\text{pr}} = k_{\text{B}} T/2$ is the average particle kinetic energy in the dimension parallel to the boundary. Hence, at the steady state,

$$T = \frac{2}{3}\frac{\overline{K}_{\text{in}}}{k_{\text{B}}} = \frac{2}{3} T_0 \tag{24}$$

which is not at thermal equilibrium (i.e., $T \neq T_0$).

Figure 12(d) depicts another 2D model that cannot relax to thermal equilibrium. The two cell borders normal to the x-axis are thermal walls. The two borders normal to the y-axis either use periodic boundary condition or are specular walls. The particle movement in the y-direction does not affect $v_{\text{x}}$. Under this condition, if particle-particle collision is ignored, the calculated steady-state $v_{\text{x}}$ would be near-zero, because $\int_0^\infty v_{\text{x}}^{-1} p_{\text{v}} dv_{\text{x}} \to \infty$. It may be understood as follows: in an ensemble of cells, the gas particles of the lowest $v_{\text{x}}$ tend to remain in the interior, while the high-$v_{\text{x}}$ particles randomly change their velocities upon colliding with the thermal walls. Eventually, every particle would have an arbitrarily small $v_{\text{x}}$, since it takes a nearly infinitely long time for such a particle to travel across the cell to collide with a thermal wall again.

It is worth noting that chaoticity may not explicitly affect the calculation of pressure. In a chaotic gas, at the container wall, the gas pressure can be obtained as

$$\overline{P}_{\text{wall}} = \frac{1}{t_{\text{n}}} \int_0^\infty \int_0^{\pi/2} (2mv \cdot \cos\theta) \cdot \overline{N}_{\text{n}}(\beta_{\text{T}} mv \cdot e^{-\beta_{\text{T}} mv^2/2}) d\theta\, dv = \rho_0 k_{\text{B}} T \tag{25}$$

with $\theta$ being the incident angle. It is the ideal gas law, as expected. In the Knudsen-gas model in Figure 12(c), without particle-particle collision, the gas pressure is

$$P_{\text{wall}} = \frac{1}{t_n} \frac{1}{2} \int_0^\infty (2mv_x)(\rho_0 v_x t_n) p_v dv_x = \rho_0 k_B T \tag{26}$$

which is also the ideal gas law. Compared to the chaotic gas, the difference in particle influx of the Knudsen gas is offset by the variation in effective temperature.

It is also worth noting that in a three-dimensional (3D) space, Figure 12(d) represents a setup with confinement in one dimension, e.g., a narrow gap between two large flat surfaces; Figure 12(a) represents a setup with confinement in two dimensions, e.g., a tube or a pore. If the setup is confined by thermal walls in all the three dimensions (e.g., a hollow cell), $p_w(v)$ should be the 3D Maxwell-Boltzmann distribution function, and $\rho_L \propto p_w/v$ can be normalized as $\rho_L(v) = m\beta_0 v \cdot e^{-\beta_0 mv^2/2}$. The effective gas-phase kinetic temperature is

$$T = \frac{2}{3k_B} \int_0^\infty \frac{mv^2}{2} \left( m\beta_0 v \cdot e^{-\beta_0 mv^2/2} \right) dv = \frac{2}{3} T_0 \tag{27}$$

In fact, if Figures 12(a) and Figure 12(d) are considered as 3D cases, with the unconfined dimensions being taken into account, $T = 2T_0/3$ generally holds true. It is compatible with Figure 12(c), in which the particle movements in different dimensions are uncorrelated.

A2. Algorithm of the computer programs

The computer programs used in the current investigation are available at [16]. In the 2D system, the particles are billiard-like finite-sized hard disks. The particle-particle and particle-wall collisions happen in the middle of the timesteps. In each timestep, the program first computes the virtual position of every particle at the end of the timestep, as if collision could not happen. Then, collision is identified as the particle-particle or particle-wall overlapping. The exact collision location and time are calculated by tracing the particle trajectories, and the correct particle information is updated by solving Newton's equations (conservation of energy and momentum). Finally, the next timestep begins. The time resolution is high, so that the expected value of the particle displacement in a timestep is less than 5% of the particle size. The probability of missing a collision is practically zero.

In Section 3, the cell-wall insertion or removal does not take time. During cell-wall insertion, if a particle overlaps with a cell wall, the particle would be moved away from the wall by one particle radius, with everything else being unchanged. If this operation conflicts with another particle, the simulation case would be abandoned.

A3. Maxwell's double-column engine

In [25], for the effect of gravity on thermal equilibrium, Maxwell pointed out: "…if the temperature of any substance, when in thermic equilibrium, is a function of the height, that of any other substance must be the same function of the height. For if not, let equal columns of the two substances be enclosed in cylinders impermeable to heat, and put in thermal communication at the bottom. If, when in thermal equilibrium, the tops of the two columns are at different temperatures, an engine might be worked by taking heat from the hotter and giving it up to the cooler, and the refuse heat would circulate round the system till it was all converted into mechanical energy, which is a contradiction to the second law of thermodynamics. The result as now given is, that temperature in gases, when in thermal equilibrium, is independent of height, and it follows from what has been said that temperature is independent of height in all other substances."

A4. Assessment of power density

To make an approximate assessment, consider the conduction electrons in a metal as a Fermi gas. Their number density is on the scale of $10^{28}$ m$^{-3}$ and the Fermi energy is on the scale of a few eV. As shown by Equation (27) in Section A1, $T \rightarrow 2T_0/3$ in a three-dimensional Knudsen gas. If this ratio is also relevant to the charge carriers and an isothermal cycle similar to Figure 8(a,b) could be designed, the energy density would be on the scale of $10^3$ J/cm$^3$. When the operation frequency is larger than 10 Hz, the upper limit of the power density may be more than 10 kW/cm$^3$.